\documentclass[final,5p,times,twocolumn]{elsarticle}

\usepackage{lineno,hyperref}

\usepackage{lipsum}
\usepackage{amsmath}
\usepackage{natbib}

\begin{document}
\begin{frontmatter}

\title{Precise Calculation of Electrical Capacitance by means of Quadruple Integrals in  Method of Moments Technique}

\author{Saeed Sarkarati\textsuperscript{a}, Mohammad Mehdi Tehranchi\textsuperscript{a,b}, Esfandiar Mehrshahi\textsuperscript{c}}
\address{\textsuperscript{a}Laser and Plasma Research Institute, Shahid Beheshti University, Tehran, Iran}
\address{\textsuperscript{b}Department of Physics , Shahid Beheshti University, Tehran, Iran}
\address{\textsuperscript{c}Department of Electrical Engineering, Shahid Beheshti University, Tehran, Iran}

\begin{abstract}
In this paper, the capacitance of a parallel plate air-gap rectangular capacitor, and a unit cube capacitor have been calculated. Because of its generality and simplicity, the method of moments (MOM) Technique is utilized. In order to improve the accuracy of the calculations, the use of quadratic integrals instead of binary integrals has been proposed. A neat form is provided for the analytical solution of the integrals required for the method of moment. The results show that there is a very small error in calculating the capacity even with coarse boundary division. The described formulas and codes can easily be used for similar purposes.
\end{abstract}


\end{frontmatter}

\section{Introduction}

Accurately calculating of electric capacitance from early water-filled Leyden jar up to modern supercapacitors has been a challenging task for scientists and researchers. \cite{Dubal2016}. The demand for fast capacitance extraction techniques increases because the usage of capacitors in sensing systems spreads widely \cite{Rahman2009}.

The desired characteristics of examined objects can be obtained by inverse problem analysis of measured parameters. In systems based on capacitive sensing such as the "capacitive imaging method", the material is scanned by two electrodes, and electrical capacitance between electrodes is recorded in each position \cite{Diamond2006, Ma2020}. And it can be used to determine the size and location of defects \cite{Liu2017}. The capacitive imaging method also can be utilized on metallic objects and is used for making a surface image of substance as an inverse problem \cite{Muttakin2018}.

To solve an "inverse problem", you first need a quick and accurate way to solve the "forward problem".  This means that the capacitance calculation must be done intelligently in a way that can be repeated without wasting CPU time. This paper focuses on extracting the capacitance of air gap capacitors. The results are utilizable to calculate the capacitance of dielectric capacitors. Also, many problems in magnetostatics can be solved similarly because of the duality of electricity and magnetism. Problems such as reconstructing the shape of defects in Magnetic flux leakage inspection can be solved in this way \cite{Dutta2008}.

The electrical capacitance of capacitors has been calculated by a variety of methods such as finite difference and finite element methods \cite{Izquierdo2009}, Monte Carlo technique \cite {LeCoZ1992, Song2020}, and method of moments \cite {Bai2004}. Except for stochastic methods, the other ones use a  matrix equation that must be solved to obtain the capacitance of desired geometry. The size of this matrix is an important parameter that affects the time required for capacitance calculation.

If we concentrate on air-gap capacitors, boundary element methods become the best choice for calculation. in  absence of dielectric medium, only air-metal boundaries must be considered. In the method of moments (MOM) the elements are placed on boundary surfaces. And in the special case of a parallel plate air-gapped capacitor, only two conducting plate surfaces are involved in generating matrix elements. So the matrix size is very small compared with the matrix size in "finite difference" and "finite element" methods. Although these two techniques create a "sparse matrix", the prolongation of the computation time due to the size of the matrix cannot be compensated for by improved "sparse matrix" manipulation techniques \cite{Liao2004, Kolundzija1998}.

MOM matrix elements can be computed by surface integration in boundary regions \cite {Tao2010}. Integral calculations can be done analytically \cite {Arcioni1997, Eibert1995, Wilton1984} or numerically \cite {Graglia1993, Rao1979}. Numerical integration can be done directly or by the Monte Carlo integration technique \cite {Mishra2009}. However, some authors have suggested a series expansion of Green's function to make the solution of the answer as a series of expressions \cite {Ayatollahi2003, Zhang2016, Hu2018, See2009}. In recent years, parallel computing methods have been used to speed up computing. \cite{Kolundzija2012, Manic2015}.

In this paper, an integral transform is used to solve quadruple integrals analytically, which givs exact formulas for MOM matrix coefficients with no approximation. These analytical expressions have been used to generate the connection matrix. So The matrix equation has been solved and electric capacitance has been calculated for parallel plate air-gap capacitor and unit cube capacitor. the results have been compared with approximated integral calculation methods. Charge distribution on capacitor plates also has been investigated.

An understandable path to solving quadruple integrals and a very neat form for analytical calculation results are provided here to help researchers calculate electrical capacitance for their own desired geometries. A Python code has been created for calculations and is now available on our website and can be used freely. NUMPY and SCIPY libraries are used to perform numerical calculations simply and relatively quickly.
\section{Method of Moments for Capacitance Calculation}

Basic ideas for MOM technique firstly suggested by great physisist James Clarke Maxwell, who wanted to calculate electric capacity of a metall square. It is worth to have a look at his work which can be found here \cite {Cavendish1967}.

He divided the square by 36 equal squares and presumed uniform charge density for each, then assumed the electric potential equal to 1 in the middle of each square piece. To keep the potential of all segments to one, the electric charge of each segment must be different from others. There are 36 values for 36 segments. but geometric symmetry shows us these 36 values can be grouped into 6 distinct values.
\begin {figure}
\center
\begin{tabular}{ c c c c c c }
  A & B & C & C & B & A \\
  B & D & E & E & D & B \\
  C & E & F & F & E & C \\
  C & E & F & F & E & C \\
  B & D & E & E & D & B \\
  A & B & C & C & B & A \\
\end{tabular}
\caption{dividing a metal square to 36 segments}
\end{figure}
He then calculated the electric charge of each segment and consequently electric capacitance of the big square.

Generally to calculate the electric capacitance of a parallel plate capacitor, at first two constant voltages, are proposed for two plates (usually 1V and -1V). Now each plate must be divided into square segments. Charge density of each segment is assumed to be constant for each segment. The electric potential of each segment must be equal to its plate. On the other hand potential of each segment can be calculated using the electric charge of all segments and coupling coefficients.
\begin{equation}
\label{eq1}
V_i = \sum_j P_{ij} Q_j
\end{equation}
This formula can be written in matrix form.
\begin{equation}
\label{eq2}
\mathbf V = \mathbf P \mathbf Q
\end{equation}
It is now necessary to solve this matrix equation to calculate the charge of each part, and it is obvious that when the electric potential and the electric charge are known, the capacitance can be obtained.
 
Coupling coefficient between two segment depends on shape and location of two segments. It can be approximated roughly by potential formula of a unit point charge. 
\begin{equation}
\label{eq3}
P_{ij} = \frac{1}{4 \pi \epsilon_0\  d_{ij}}
\end{equation}
where $d_{ij}$ is the distance between centers of two segments.

In literature, this approximation is used in a method called "surface charge simulation method"\cite {Zhou1993} although this method can be categorized in varieties of MOM. This formula is not useful for obtaining the self-coupling coefficient. Self-coupling is calculated by integrating over the area to find the mean distance of all the points of area to its center.
\begin{equation}
\label{eq4}
P_{ii} = \frac 1 {S_i} \int_{x \in S_i} \int_{y \in S_i} \frac {dx\  dy}{4 \pi \epsilon_0 \sqrt{(x-x_{ci})^2 + (y-y_{ci})^2}}
\end{equation}
Where $(x_{ci}, y_{ci})$ is the center point of domain i and $S_i $ is the segment area.
For less approximate results in calculating the capacity, it is better to use a double integral formula, not only for the self-coupling but also for the mutual coupling. It leads us to double integral formulation.
\begin{equation}
\label{eq5}
P_{ij} = \frac 1 {S_i}\int_{x \in S_i} \int_{y \in S_i} \frac {dx\  dy}{4 \pi \epsilon_0 \sqrt{(x-x_{cj})^2 + (y-y_{cj})^2+ z^2}}
\end{equation}
Two domains are assumed to be parallel. $z$ is the distance of two domains and is constant over integration. The integral must be calculated over domain i, and the distance of each point to the center of other domains is considered in this formula. These integrals can be calculated analytically and have been applied to find the capacitance of parallel plate capacitors by Nishiyama and Nakamura\cite {Nishiyama1994}.

Obviously this formula is not suitable where two domains are relatively close according to their dimensions. In this case, the center of one domain can not be proposed as representative of all points. Really it is better to find all mutual distances between points of two segments. It can be done by using quadruple integral instead of double integral.
\begin{equation}
\label {eq6}
P_{ij} = \frac 1 {S_i S_j}\iint\limits_{x_i, y_i \in S_i}\ \  \iint\limits_{x_j, y_j \in S_j}  \frac{dx_i dy_i dx_j dy_j}{4 \pi \epsilon_0 d_{ij}}
\end{equation} 
where $d _{ij}$ is 
\begin{equation}
\label {eq7}
d_{ij} = \sqrt{(x_i-x_j)^2 + (y_i-y_j)^2 + z^2}
\end{equation} 
\section{Quadruple Integration for Parallel Rectangular Segments}
To obtain coupling coefficients described in Eq(6) one can use the numerical integration method or solve it analytically. Although there are many improvements in numerical techniques, analytical solutions still have less process time. Firstly it has been done by Eibert and Hansen for triangular domains\cite{Eibert1995}. Analytical solution for rectangular domains presented by Lopez-Pena and Mosig \cite{Lopez2009}, with a small mistake in the derived formula. Recently this integral has been performed by Maccarrone and Paffuti\cite{Maccarrone2017}, and its result has been used to find capacitance and forces for two square electrodes.These integrals have been calculated by Zhenfei Song et al \cite{Song2011} for calculating partial inductance.

To perform integral in Equation(6) this integral transformation can be used\cite{Ciftja2010}.
\begin{equation}\label {eq8}
\frac 1 d_{ij} = \frac{2}{\sqrt{\pi}}\int_0^{\infty} e^{-u^2 d_{ij}^2 }du
\end{equation}
And Equation(6) can be rewritten in this form.
\begin{equation}
\label {eq9}
P_{ij} = \frac 2 {\sqrt{\pi} S_i S_j}\int_0^{\infty}\iint\limits_{x_i, y_i \in S_i}\ \  \iint\limits_{x_j, y_j \in S_j} \\ \frac{e^{-u^2 d_{ij}^2 } dx_i dy_i dx_j dy_j}{4 \pi \epsilon_0 } du
\end{equation}

For simplicity, we remove constants from the formula.

\begin{equation}
\label {eq10}
I = \int_0^{\infty}\iint\limits_{x_i, y_i \in S_i}\ \  \iint\limits_{x_j, y_j \in S_j}  e^{-u^2 d_{ij}^2} dx_i dy_i dx_j dy_j du
\end{equation}

There is no way to find a primitive function of $e^{-u^2 d_{ij}^2}$ over these five integrals, but primitive function over four inner integrals can be found. Assume J as the primitive function of the quadruple integral.
\begin{equation}
\label {eq11}
J = \int \int \int \int e^{-u^2 d_{ij}^2 } dx_i dy_i dx_j dy_j  
\end{equation}
Then J can be calculated analytically. It can be done by common mathematical softwares. The solution has been obtained by use of "Wolfram Alpha" \cite{Wolfram}.

\begin{equation}\label {eq12}
\begin{split}
j = \frac{e^{-u^2(x^2+y^2+z^2)}}{4u^4} + \frac{\sqrt{\pi}e^{-u^2(y^2 + z^2)}x\; \mathrm{erf}(u x)}{4u^3} \\
+\frac{\sqrt{\pi}e^{-u^2(x^2 + z^2)}y \mathrm {erf}(u y)}{4u^3} + \frac{\pi e^{-u^2 z^2}x\;y\; \mathrm{erf}(u x)\; \mathrm{erf}(u y)}{4u^2}
\end{split}
\end{equation}
Where $x = x_i - x_j$ , $y = y_i - y_j$ and $z = z_c$\ . 
Now, these integrals must be calculated separately.
\begin{equation}\label {eq13}
\begin{split}
I_1\; = \;&\bigg[\int_0^{\infty}\frac{e^{-u^2(x^2+y^2+z^2)}} {4u^4}\bigg]_{D_0}^{D_1}\\
I_2\; = \;&\bigg[\int_0^{\infty}\frac{\sqrt{\pi}e^{-u^2(y^2 + z^2)}x\; \mathrm{erf}(u x)}{4u^3}\bigg]_{D_0}^{D_1}\\
I_3\; = \;&\bigg[\int_0^{\infty}\frac{\sqrt{\pi}e^{-u^2(x^2 + z^2)}y\; \mathrm {erf}(u y)}{4u^3}\bigg]_{D_0}^{D_1}\\
I_4\; = \;&\bigg[\int_0^{\infty}\frac{\pi e^{-u^2 z^2}x\; y\; \mathrm{erf}(u x) \mathrm{erf}(u y)}{4u^2}\bigg]_{D_0}^{D_1}
\end{split}
\end{equation}

We want to do calculations over rectangular segments. So the limits of integration are over two rectangles.

\begin{equation}\label {eq14}
\begin{split}
D_0:\;\;x_i=a_0, y_i = b_0,\ x_j = c_0,y_j = d_0\\
D_1:\;\;x_i=a_1, y_i = b_1, \ x_j = c_1,y_j = d_1
\end{split}
\end{equation}

Finally, the answer can be calculated.

\begin{equation}\label {eq15}
\begin{split}
I_1 = &\sum_{i,j,k,l=0}^1 A_{i,j,k,l}\frac {\sqrt{\pi} } {6} (x^2+y^2+z^2)^{\frac 3 2}\\
I_2 = &\sum_{i,j,k,l=0}^1 -A_{i,j,k,l}\frac {\sqrt{\pi} } {4}\;x\ \times \\
&\Bigg((y^2 + z^2) \;\mathrm{sinh^{-1}}(\frac{x}{\sqrt{y^2 + z^2}}) +x \sqrt{x^2 + y^2 + z^2}\Bigg)\\ 
I_3 = &\sum_{i,j,k,l=0}^1 -A_{i,j,k,l}\frac {\sqrt{\pi} } {4}\;y\ \times \\
&\Bigg((x^2 + z^2)\;\mathrm{sinh^{-1}}(\frac{y}{\sqrt{x^2 + z^2}}) +y \sqrt{x^2 + y^2 + z^2}\Bigg) \\ 
I_4 = &\sum_{i,j,k,l=0}^1 A_{i,j,k,l}\frac {\sqrt{\pi} } {2}\;x\;y\; \Bigg(x\;\mathrm{sinh^{-1}}(\frac{y}{\sqrt{x^2 + z^2}})+ \\
 &y\;\mathrm{sinh^{-1}}(\frac{x}{\sqrt{y^2 + z^2}}) -z\;\mathrm{tan^{-1}}(\frac{x\;y}{z\;\sqrt{x^2+y^2 + z^2}}) \Bigg)\\ 
\end{split}
\end{equation}

In above expressions $x = a_i - c_j$ , $y = b_k - d_l\ $ and $z = z_c$. The amount of $A_{i,j,k,l}$ deponds on sum of i to l. If this summation is an odd number $A_{i,j,k,l}$ becomes -1, otherwise its value equals to 1.
\begin{equation}\label{eq16}
A_{i,j,k,l} = \Big\{^{\;\;1\quad  \mathrm{if}\; i + j + k + l\; \mathrm{is \;even}}_{-1\quad  \mathrm{if}\; i + j + k + l\; \mathrm{is \;odd}}
\end{equation}

Now it is needed to find the sum of $I_1$ to $I_4$.

\begin{equation}\label {eq:17}
\begin{split}
I&=I_1+I_2+I_3+I_4 = \sum_{i,j,k,l=0}^1 A_{i,j,k,l}\\
&  \bigg[ \frac {\sqrt{\pi} } {12} \Big((-x^2-y^2+2\;z^2)\sqrt{x^2 + y^2 + z^2} \Big) \\
& \ +  \frac {\sqrt{\pi} } {4} \Big( y(x^2 - z^2)\;\mathrm{sinh^{-1}}(\frac{y}{\sqrt{x^2 + z^2}}) \Big)\\
&\ +   \frac {\sqrt{\pi} } {4}\Big( x(y^2-z^2)\;\mathrm{sinh^{-1}}(\frac{x}{\sqrt{y^2 + z^2}})  \Big)\\ 
&\ -  \frac {\sqrt{\pi} } {2}x \;y\;z\;\mathrm{tan^{-1}}(\frac{x\;y}{z\;\sqrt{x^2+y^2 + z^2}})\bigg]
\end{split}
\end{equation}

As is known $"\mathrm{sinh^{-1}}(x) = \mathrm{ln}(x+\sqrt{x^2 + 1})"$,
so the above formula can be rewritten in this form.

\begin{equation}\label {eq:18}
\begin{split}
I&=\sum_{i,j,k,l=0}^1 A_{i,j,k,l}\bigg[\frac {\sqrt{\pi} } {12} (-x^2-y^2+2\;z^2)\sqrt{x^2 + y^2 + z^2}   \\
& + \frac {\sqrt{\pi} } {4}  y\;(x^2 - z^2)\;\mathrm{ln}\Big(\frac{y+\sqrt{ {x^2 + y^2 + z^2}}}{\sqrt{x^2 + z^2}}\;\Big) \\
& + \frac {\sqrt{\pi} } {4} x\;(y^2-z^2)\; \mathrm{ln}\Big(\frac{x+\sqrt{ {x^2 + y^2 + z^2}}}{\sqrt{y^2 + z^2}}\;  \Big)\\ 
&- \frac {\sqrt{\pi} } {2}x \;y\;z\;\mathrm{tan^{-1}}(\frac{x\;y}{z\;\sqrt{x^2+y^2 + z^2}} ) \bigg]
\end{split}
\end{equation}

In this paper, we prefer to use the hyperbolic form of formula, which introduced in eq(17).
In the case that two segments are coplanar, mutual coupling obtains by taking the limit of equation(17) when $z$ goes to zero.
\begin{equation}\label {eq:19}
\begin{split}
I_{coplanar}=\sum_{i,j,k,l=0}^1  A_{i,j,k,l} \bigg[  \frac {\sqrt{\pi} } {12} \Big((-x^2-y^2)\sqrt{x^2 + y^2} \Big)\ + \\
 \frac {\sqrt{\pi} } {4} \Big( y(x^2 )\;\mathrm{sinh^{-1}}(\frac{y}{x })+x(y^2)\;\mathrm{sinh^{-1}}(\frac{x}{y}) \Big) \bigg]
\end{split}
\end{equation}

Implementation of this formula needs special attention to the cases in which any denominator of fractions becomes zero. Easily each term, involving such fractions goes to zero and can be omitted in calculations.

And finally, the self-coupling of a rectangular segment takes this form.

\begin{equation}\label {eq:20}
\begin {split}
I_{SC} = \frac {\sqrt{\pi} } {3} (x^3+y^3) 
+ \frac {\sqrt{\pi} } {3} \left((-x^2-y^2) \sqrt{x^2 + y^2}\ \right) \\ +\sqrt{\pi} \left( y(x^2 )\;\mathrm{sinh^{-1}}(\frac{y}{x})+x(y^2)\;\mathrm{sinh^{-1}}(\frac{x}{y})  \right)
\end {split}
\end{equation}
Now x and y are length and width of rectangular region.

\section{Quadruple Integration for Perpendicular Rectangular Segments}
Coupling coeficient for perpendicular segments can be obtained by this formula.

\begin{equation}
\label {eq21}
P_{ij} = \frac 1 {S_i S_j}\int \int \int \int \frac{dx_i dy_i dx_j dz_j}{4 \pi \epsilon_0 d_{ij}}
\end{equation} 
where $d _{ij}$ is 
\begin{equation}
\label {eq22}
d_{ij} = \sqrt{(x_i-x_j)^2 + (y_i-y_c)^2 + (z_c - z_j)^2}
\end{equation} 
Where $z_c$ and $y_c$ are constants in integration. using integral transform utilized for parallel segments, primitive function of quadraple integral can be obtained.
\begin{equation}
\label {eq23}
\begin{split}
J = &\frac{\pi\sqrt{\pi}\;x\; \mathrm {erf}(u x)\; \mathrm {erf}(u y)\; \mathrm {erf}(u z)}{8u^3}\\
+\; &\frac{\pi e^{-u^2 x^2} \mathrm{erf}(u y)\; \mathrm{erf}(u z)}{8u^4}
\end{split}
\end{equation}
Where $x = x_i - x_j$ , $y = y - y_j$ and $z = x_i - x_j$\ .
And finally coupling coefficient of two perpendicular plate is obtained.

\begin{equation}\label {eq:24}
\begin{split}
I& = \sum_{i,j,k,l=0}^1 A_{i,j,k,l}\\
&  \bigg[ -\frac {\sqrt{\pi} } {6} \left((y\; z)\sqrt{x^2 + y^2 + z^2} \right) \\
& +\  \frac {\sqrt{\pi} } {12} \bigg( z(3x^2 - z^2)\;\mathrm{sinh^{-1}}(\frac{y}{\sqrt{x^2 + z^2}}) \bigg)\\
& +\  \frac {\sqrt{\pi} } {12} \bigg( y(3x^2 - y^2)\;\mathrm{sinh^{-1}}(\frac{z}{\sqrt{x^2 + y^2}}) \bigg)\\
& +\  \frac {\sqrt{\pi} } {2} \bigg( x\; y\; z\;\mathrm{sinh^{-1}}(\frac{x}{\sqrt{y^2 + z^2}}) \bigg)\\
&-  \frac {\sqrt{\pi} } {4}\;x \;z^2\;\mathrm{tan^{-1}}(\frac{x\;y}{z\;\sqrt{x^2+y^2 + z^2}})\\
&-  \frac {\sqrt{\pi} } {4}\;x \;y^2\;\mathrm{tan^{-1}}(\frac{x\;z}{y\;\sqrt{x^2+y^2 + z^2}})\\
&-  \frac {\sqrt{\pi} } {12}\;x^3\;\mathrm{tan^{-1}}(\frac{y\;z}{x\;\sqrt{x^2+y^2 + z^2}})\bigg]
\end{split}
\end{equation}

Now in above expressions $x = a_i - c_j$ , $y = b_k - y_c\ $ and $z = z_c - d_l$.
\section{Results}
\begin {figure}[h]
	\center
	\includegraphics[width=\linewidth]	{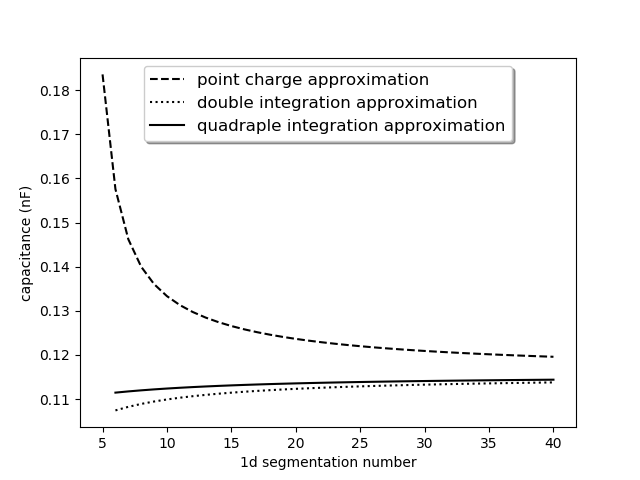}
	\caption{Capacitane of parallel plate capacitor with three  methods}
\end{figure}
To test the analytical results, they are first compared with the numerical integration results. A complete agreement and zero error occurred. Although because of singularities, numerical integration can not be done easily for every domain of integration. For testing the ability of formulas, two classical problems were solved with these formulas: Capacitance of parallel plate air-gap capacitor and Capacitance of unit cube.

The capacitance of the parallel plate air-gap capacitor has been calculated in three methods. In the first one, point charge approximation is used for mutual-coupling and double integration for self-coupling. In the second method, both self and mutual couplings are calculated through double integration. The third method uses quadruple integration for calculating coupling coefficients. 

\begin {figure}[h]
	\center
	\includegraphics[width=\linewidth]	{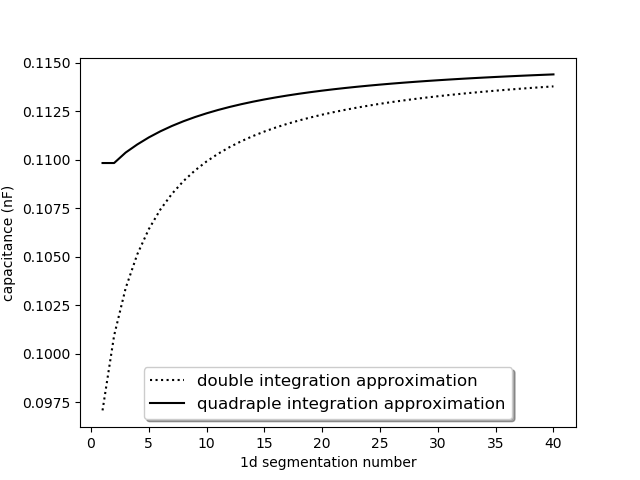}
	\caption{comparsion between results extracted from double and quadraple integrations in calculating capacitance of parallel plate capacitor}
\end{figure}
A capacitor with dimensions of $1m \times 1m $ for plates and 10 cm for the separation gap is considered.In fig.2 the results of capacitance extraction are shown for these three methods versus the number of segmentation of square plate in each dimension (n). The total number of tiles is 2 * n * n. Due to the very inaccurate answers of the point charge approximation method in coarse segmentation, the first five results are omitted. It can be seen that point charge approximation is out of accuracy and two others are more compatible.

In fig.3 double and quadruple integration methods are compared. Clearly quadruple integration results in more accurate advantages. Even in coarse segmentation, a valid answer has been achieved from quadruple integration.

\begin {figure}[]
	\center
	\includegraphics[width=\linewidth]	{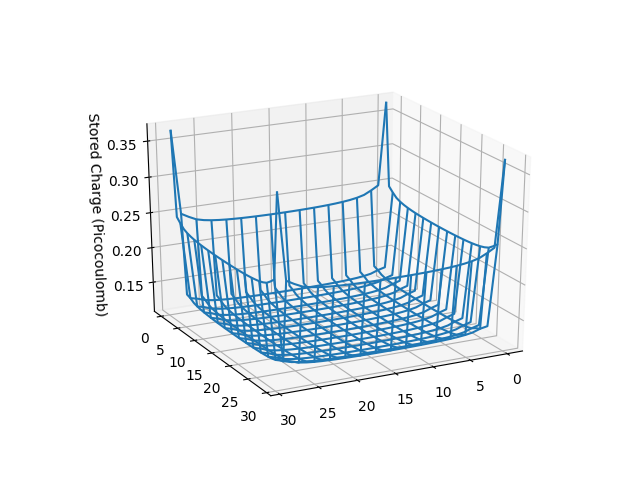}
	\caption{Charge distribution on one plate of parallel plate plate air gapped capacitor}
\end{figure}
\begin {figure}[]
	\center
	\includegraphics[width=\linewidth]	{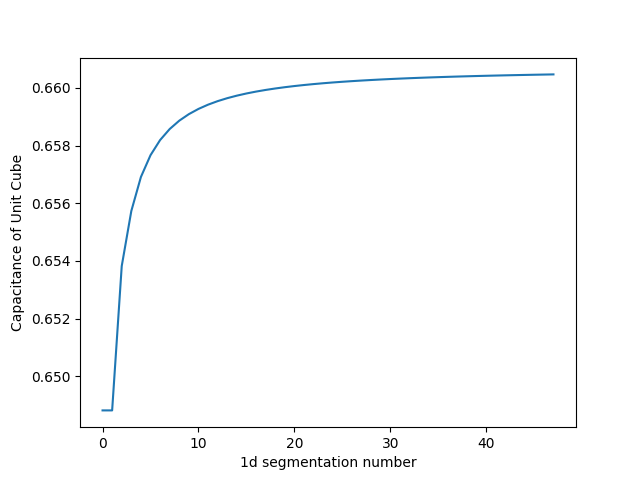}
	\caption{Capacitance of unit cube in units of $4\pi \epsilon_0$}
\end{figure}
As mentioned above, in MOM a set of equations is solved to obtain the charge amount of every tile. So charge density distribution is obtained directly in this method. As an example, this method applied for an air-gap parallel plate capacitor with dimensions of $1m \times 1m $ for plates and 10 cm for the separation gap. In fig. 2 charge distribution on the upper plate has been shown. Calculating charge density in some research areas, such as high voltage engineering, is a bottleneck in apparatus design.

The last investigated problem is capacitance of unit cube. It is a classic problem in electrostatics which can not be solved exactly. Many aouthors have tried different methods to solve this problem since 1950\cite{Reitan1951}. We tried our analytical results for extracting matrix elements needed for method of moments. this results in 73.385 pF for 48 segment in each dimension i.e. 48 * 48 * 6 tiles totally. Most of the authors report this amount in units of $4\pi \epsilon_0$. It means that this amount must be multiplied by $9 \times 10^9$ and results in 0.66047, which is very close to best claimed amount i.e. 0.660678\cite{Hwang2010}. The amount of capacitance have been drawn versus 1d segmentation number in fig.5.

This chart shows not only good results in fine segmentation but also acceptable results in coarse segmentation. even if we consider each face of the cube as one segment, it yields less than 2 pecent error.
\section{conclusion}
In this paper, we have introduced neat formulas for extracting coupling coefficients in the MOM connection matrix. We have used these formulas for obtaining coupling coefficients. Then we used MOM to obtain electric capacitance of parallel air-gapped capacitor and capacitance of the unit cube. Electric capacitance versus the number of segmentation is investigated. We showed that even in very coarse segmentation of boundaries, very good results can be obtained from MOM. So it can be used for achieving the solution of the "forward problem" when you need a quick and accurate method.
\section*{References}
\bibliography{sarkarati.bib}
\end{document}